\documentclass[aps,pra,onecolumn,notitlepage,superscriptaddress,showpacs,longbibliography,10pt]{revtex4-1}

\usepackage{amsmath}    
\usepackage{amsfonts}
\usepackage{bm}
\usepackage{amssymb}
\usepackage{nicefrac}
\usepackage{appendix}
\usepackage{graphicx}   
\usepackage{subfigure}  
\usepackage{color}
\usepackage{siunitx}
\usepackage{listings}
\usepackage{hyperref}
\hypersetup{
    colorlinks=true,
    linkcolor=blue,
    urlcolor=blue,
}
\renewcommand{\eqref}[1]{Eq.~(\ref{eq:#1})}
\newcommand{\Eqref}[1]{Equation~(\ref{eq:#1})}
\newcommand{\eqrefs}[2]{Eqs.~(\ref{eq:#1}--\ref{eq:#2})}
\newcommand{\Eqrefs}[2]{Equations~(\ref{eq:#1}--\ref{eq:#2})}

\newcommand{\re}{\textrm{Re}}
\newcommand{\im}{\textrm{Im}}

\newcommand{\GAMMA}{\boldsymbol{\Gamma}}

\renewcommand{\vec}[1]{\mathbf{#1}}
\newcommand{\defn}{\stackrel{\mathrm{def}}{=}}

\begin{document}

\title{Modeling lasers and saturable absorbers via multilevel atomic media in the \textit{Meep}~FDTD software: Theory and implementation}
\author{Alexander Cerjan}
\affiliation{Department of Physics, Pennsylvania State University, University Park PA 16802}
\author{Ardavan Oskooi}
\affiliation{Simpetus LLC, San Francisco CA 94109}
\author{Song-Liang Chua}
\affiliation{DSO National Laboratories, Singapore}
\author{Steven G. Johnson}
\affiliation{Department of Mathematics, Massachusetts Institute of Technology, Cambridge MA 02139}

\date{\today}

\begin{abstract}
This technical note describes the physical model, numerical implementation, and validation of multilevel atomic media for lasers and saturable absorbers in \emph{Meep}: a free/open-source finite-difference time-domain (FDTD) software package for electromagnetics simulation.  Simulating multilevel media in the time domain involves coupling rate equations for the populations of electronic energy levels with Maxwell's equations via a generalization of the Maxwell--Bloch equations. We describe the underlying equations and their implementation using a second-order discretization scheme, and also demonstrate their equivalence to a quantum density-matrix model.  The \emph{Meep} implementation is validated using a separate FDTD density-matrix model as well as a frequency-domain solver based on steady-state \textit{ab-initio} laser theory (SALT).
\end{abstract}

\maketitle

\section{Introduction}

One of the primary methods in computational electromagnetics is finite-difference
time-domain (FDTD), which solves Maxwell's equations using a discretized temporal and spatial grid~\cite{taflove_computational_2005,taflove13}.
Since its public release in 2006, the open source FDTD software package \emph{Meep} (\href{https://meep.readthedocs.io/en/latest/}{meep.readthedocs.io}) has become a widely used tool for photonics research and development~\cite{oskooi_meep:_2010}. In its original implementation, \emph{Meep} was limited to
simulating either linear dielectric or conductive media, or instantaneous versions of the Pockels and Kerr effects.
However, many important optical phenomena involve saturable nonlinear media, such as
optical bistability and lasing, which has not been supported.

In this technical note, we outline \emph{Meep}'s recent addition of saturable nonlinear media which enables the simulation of lasers and saturable absorbers.  This is achieved by coupling Maxwell's equations (for the electromagnetic fields) to rate equations for atomic level populations via a polarization field, and co-evolving the resulting FDTD-discretized ``Maxwell--Bloch'' equations~\cite{taflove_computational_2005,Nagra1998,Jiang2000,Chang2004,Huang2006,Bermel2006,boehringer_full_2008,Chua2011}. This implementation
is then validated against simulations of stable, multimode lasing calculated using the steady-state
\textit{ab-initio} laser theory (SALT) \cite{tureci_self-consistent_2006,tureci_strong_2008,ge_steady-state_2010,esterhazy_scalable_2014}. We also discuss how to derive the classical oscillator
equations for a saturable nonlinear medium from the quantum mechanical equations of motion for
an electron bound to an atom or molecule interacting with an electric field, i.e.\ the Bloch equations.
Finally, we review different unit conventions in the literature.

\section{Evolution equations for saturable media}

In saturable nonlinear media, the medium possesses a set of internal degrees of freedom, the occupations of its electronic states,
that are coupled to the electromagnetic field through the dipole moments of possible transitions between these states.
Mathematically, this coupling between the electric field and the saturable medium can be modeled either by using
a classical oscillator equation for the polarization of the saturable medium, or via the evolution of the density matrix of the
electronic states. In this section, we will first discuss the classical oscillator model, as these are the equations used
in \emph{Meep} for saturable media. Then, we will derive the equivalence between the classical oscillator equation
and the density-matrix evolution equations.

\subsection{Classical oscillator model \label{sec:osc}}

To model saturable nonlinear media, \emph{Meep} uses a classical oscillator equation
for the nonlinear portion of the polarization of the media,
\begin{equation}
  \frac{d^2\vec{P}_n}{dt^2} + \gamma_n \frac{d\vec{P}_n}{dt} + \left(\omega_n^2 + \left(\frac{\gamma_n}{2} \right)^2 \right) \vec{P}_n = -\Delta N_n(\vec{x},t) \bar{\sigma}_n \vec{E}(\vec{x},t), \label{eq:oscP}
\end{equation}
coupled to a set of rate equations for the evolution of the populations within each electronic state,
\begin{equation}
  \frac{\partial N_i(\vec{x})}{\partial t} = - \sum_j \Gamma_{ij} N_i(\vec{x}) + \sum_j \Gamma_{ji} N_j(\vec{x}) + \sum_n \Xi_{i,n} \left[\frac{1}{\omega_n \hbar} \vec{E}(\vec{x},t) \cdot \left(\frac{\partial }{\partial t} + \frac{\gamma_n}{2} \right) \vec{P}_n(\vec{x},t) \right]. \label{eq:oscN}
\end{equation}
There are many subtle details about these equations, but let us first review the notation used in these equations.
\begin{align}
  &\vec{E}(\vec{x},t) \in \mathbb{R} \textrm{ \textemdash electric field vector} \notag \\
  &\vec{P}_n(\vec{x},t) \in \mathbb{R} \textrm{ \textemdash nonlinear polarization density vector of the $n$th electronic transition} \notag \\
  &N_i(\vec{x},t) \in \mathbb{R} \textrm{ \textemdash population density of the $i$th electronic state across all atoms/molecules at $\vec{x}$} \notag \\
  &\Delta N_n(\vec{x},t) = N_i(\vec{x},t) - N_j(\vec{x},t) \textrm{ \textemdash inversion of the population of the $n$th dipole transition} \notag \\
  &\omega_n \in \mathbb{R} \textrm{ \textemdash central transition frequency of the $n$th electronic transition} \notag \\
  &\gamma_n \in \mathbb{R} \textrm{ \textemdash full-width half-maximum linewidth of the $n$th transition} \notag \\
  &\Gamma_{ij} \in \mathbb{R} \textrm{ \textemdash non-radiative decay/pumping rate from state $i$ to state $j$} \notag \\
  &\bar{\sigma}_n \in \mathbb{R}^{(3 \times 3)} \textrm{ \textemdash coupling tensor between the electric field and nonlinear polarization} \notag \\
  &\Xi_{i,n} = 0, \pm 1 \textrm{ \textemdash only non-zero if state $i$ is the upper ($+1$) or lower ($-1$) state in transition $n$} \notag
\end{align}
In \eqrefs{oscP}{oscN}, there are two separate sets of indices, $i,j$ which correspond to electronic states of the saturable medium, while $n$ denotes
a transition between two of these states that can potentially interact with the electric field. 
If $E_i$ and $E_j$ are the energies of these two electronic states, the corresponding central transition
frequency is $\omega_n = (1/\hbar)(E_i - E_j)$.
The two states linked by a transition are said to be `inverted' if $N_i > N_j$ and $E_i > E_j$, i.e.\ that the higher energy
state has a greater occupation than the lower energy state. In this case, the transition will yield spontaneous and stimulated emission.
If the medium is not inverted, the transition will instead act as an absorber for incident electric fields with frequencies similar to the
transition frequency, $\omega_n$.
In total, there are $M$ partial differential equations which describe the evolution of the occupation of each of the electronic states, and
$N$ oscillator equations for each of the different nonlinear polarization fields.

In principle, every $i \ne j$ pair of states could have a non-zero dipole moment yielding a nonlinear polarization field, $\vec{P}_n$.
However, in practice, many of these potential transitions can be ignored, either because the dipole matrix element is zero, meaning that $\bar{\sigma}_n = 0$, or because the two states
have low occupations, usually because the non-radiative decay rates out of these states are much faster than the rate of stimulated emission, such that $\Delta N_n \approx 0$.
This second condition can also occasionally occur because the two states are always approximately equally populated.
In writing \eqref{oscN} the only transition terms (those in the square brackets) which are included in the evolution of electronic state $i$ are those
which actually couple to state $i$, such that $\Xi_{i,n} = \pm 1$, otherwise $\Xi_{i,n} = 0$.

The tensor $\bar{\sigma}_n$ represents the effective coupling strength between the nonlinear polarization field and the
electric field. One can understand that this quantity must be a tensor because the dipole element between two electronic states (a vector),
\begin{equation}
  \boldsymbol{\theta}_n = e \langle \psi_i | \hat{\vec{x}} | \psi_j \rangle,
\end{equation}
in which $|\psi_i \rangle$ is the wavefunction of the $i$th electronic state and $e$ is the charge of an electron,
is not, in general, parallel to the electric field, while the induced nonlinear polarization is necessarily
parallel to $\boldsymbol{\theta}_n$. Thus, it can be shown that
\begin{equation}
  \bar{\sigma}_n = \left(\frac{2 \omega_n}{\hbar} \right) \boldsymbol{\theta}_n^* \otimes \boldsymbol{\theta}_n.
\end{equation}
In practice, $\bar{\sigma}_n$ can often be treated as a scalar, $\sigma_n = 2\omega_n|\theta_n|^2/\hbar$,
as one is typically interested in understanding the response of a nonlinear
medium in the regime where it maximally interacts with the electric field, i.e.\ when $\boldsymbol{\theta}_n \parallel \vec{E}$.

Finally, there are two additional terms in \eqrefs{oscP}{oscN} that are atypical when writing
a classical oscillator model and associated population rate equations, $(\gamma_n/2)^2 \vec{P}_n$ on the left side
of \eqref{oscP}, and $(\gamma_n/2)\vec{P}_n$ on the right side of \eqref{oscN}. Typically, these terms
are approximated to zero on the assumption that $\omega_n \gg \gamma_n$. However, as will be shown in Sec.\ \ref{sec:comp},
these terms are necessary to find proper agreement with a density matrix model of saturable media, and so we include them here.

\subsection{Implementation in \emph{Meep} \label{sec:impl}}

\Eqrefs{oscP}{oscN} are implemented in the \emph{Meep} FDTD code via second-order-accurate centered-difference approximations.  In such a discretization scheme, the key question is what spatial and temporal sampling points are used for each variable.  The electric field $\vec{E}$ in \emph{Meep} is discretized on a Yee grid~\cite{taflove_computational_2005}, with the $k$-th component $E_k$ sampled at points
$$
E_{k,\vec{i}+\vec{e}_k/2}^m \defn E_k(m\Delta t, (\vec{i}+\vec{e}_k/2)\Delta x) \, ,
$$
where $\vec{i} = (i_1,i_2,i_3) \in \mathbb{Z}^3$ is an integer coordinate of a grid point (Yee voxel vertex) and $\vec{e}_k$ is the Cartesian unit vector in direction $k$.   Compared to this grid, the electric polarization-density components $P_{n,k}$ are sampled at the \emph{same} Yee grid points $P_{n,k,\vec{i}+\vec{e}_k/2}^m$.  In contrast, the population-density components $N_i$ are sampled as $N_{i,\vec{i}+\vec{o}/2}^m$ at the \emph{center} $\vec{i}+\vec{o}/2$ of each Yee voxel, where $\vec{o}=(1,1,1)$, at the same time instant $m\Delta t$.   This leads to the following equations to update $\vec{P}_n$ and $N_i$ at each timestep $m$ of the FDTD simulations: given $\{N_i^{m-1}, \vec{P}_n^{m-1},\vec{P}_n^{m},\vec{E}^{m-1}\}$, \emph{Meep} computes $\{\vec{E}^{m}$, $N_i^{m},\vec{P}_n^{m+1}\}$ (in that order).   The timestepping of the electromagnetic fields via Maxwell's curl equations~\cite{taflove_computational_2005} is essentially unmodified except for the coupling $\vec{E} = \vec{D}-\sum_n\vec{P}_n$ to the polarization field.  $\vec{E}^{m}$ is updated first because it depends only on $\vec{P}_n^{m}$; then $N_i^{m}$ is updated next because it depends on $\{N_j^{m-1},\vec{E}^{m},\vec{E}^{m-1},\vec{P}_n^{m},\vec{P}_n^{m-1}\}$ as described below in \eqref{oscN-Meep}; then $\vec{P}_n^{m+1}$ is computed from $\{N_i^{m}, \vec{P}_n^{m-1},\vec{P}_n^{m},\vec{E}^{m}\}$ via \eqref{oscP-Meep} below.  (In consequence, \emph{Meep} must store $\vec{P}_n$ and $\vec{E}$ from two consecutive timesteps, and can otherwise update its data in-place.)

First, we update $N_j$ for each level $j=1,\ldots,L$ to obtain $N_j^m$ from $N_j^{m-1}$.
Let $\vec{N} \defn (N_1, \ldots, N_L)$ denote the vector of population densities for all $L$ levels being tracked in \eqref{oscN}, and let $\GAMMA$ denote the $L \times L$ matrix of the $\Gamma_{ij}$ transition rates.  Let $ w_j^{m-\nicefrac{1}{2}}$ denote the rate of work being done on (or by) level $j$ in \eqref{oscN} evaluated at time $(m-\nicefrac{1}{2})\Delta t$ for each voxel center $\vec{i}+\vec{o}/2$:
\begin{equation}
    w_{j,\vec{i}+\vec{o}/2}^{m-\nicefrac{1}{2}} \defn \sum_n \frac{\Xi_{i,n}}{\omega_n \hbar} \overline{\left[ \frac{\vec{E}^{m-1} + \vec{E}^{m}}{2} \cdot \left(\frac{\vec{P}_n^{m} - \vec{P}_n^{m-1}}{\Delta t}+ \frac{\gamma_n}{2}\frac{\vec{P}_n^{m-1} + \vec{P}_n^{m}}{2}   \right)  \right]}_{\vec{i}+\vec{o}/2}.
\end{equation}
Note that we must average timesteps $m$ and $m-1$ to obtain the electric and polarization fields at timestep $m-\nicefrac{1}{2}$ to second-order accuracy, while $\partial/\partial t$ is also computed by a second-order center-difference approximation around $m-\nicefrac{1}{2}$.  The $\overline{[\cdots]}_{\vec{i}+\vec{o}/2}$ term denotes the spatial averaging required to compute the dot product at the voxel center $\vec{i}+\vec{o}/2$ to second-order accuracy.    That is, for each component $k$ we first compute the corresponding dot-product term $E_k P_{n,k}$ at the Yee grid point $\vec{i}+\vec{e}_k$, and then we average four neighboring Yee-grid points to obtain the value at the center of the voxel.   Let $\vec{w} \defn (w_1, \ldots, w_L)$ denote a vector of these work values.   The second-order discretized \eqref{oscN} for $\vec{N}$ is then
\begin{equation}
   \left[\frac{\vec{N}^m - \vec{N}^{m-1}}{\Delta t}\right]_{\vec{i}+\vec{o}/2} = (\GAMMA^T - \GAMMA) \left[\frac{\vec{N}^m + \vec{N}^{m-1}}{2} + \vec{w}^{m-\nicefrac{1}{2}}\right]_{\vec{i}+\vec{o}/2} 
  \label{eq:oscN-Meep}
\end{equation}
(resembling a Crank-Nicolson scheme).  \Eqref{oscN-Meep} can be rearranged to solve for $\vec{N}^m$.   (Note that this rearrangement will include the matrix $[I - \frac{\Delta t}{2} (\GAMMA^T - \GAMMA)]^{-1}$, which is precomputed before timestepping begins.)

Second, we update $\vec{P}_n$ to compute $\vec{P}_n^{m+1}$. From \eqref{oscP}, we obtain the following center-difference discretization (supporting diagonal $\bar{\sigma}_{n}$ matrices):
\begin{equation}
\left[ \frac{P_{n,k}^{m+1} - 2P_{n,k}^{m} + P_{n,k}^{m-1}}{\Delta t^2} + \gamma_n \frac{P_{n,k}^{m+1} -P_{n,k}^{m-1}}{2\Delta t} + \left( \omega_n^2 + \left(\frac{\gamma_n}{2}\right)^2\right) P_{n,k}^{m}     \right]_{\vec{i}+\vec{e}_k/2} = - \left[ \overline{\Delta N_n}^m \bar{\sigma}_{n,k,k} E_{k}^{m} \right]_{\vec{i}+\vec{e}_k/2} \, ,
\label{eq:oscP-Meep}
\end{equation}
where $\overline{\Delta N_{n}}^m$ denotes the population density difference (between the $n$-th transition's levels) averaged over the four neighboring voxel centers to obtain the value of $\Delta N_n$ at $\vec{i}+\vec{e}_k/2$ to second-order accuracy in $\Delta x$.  \Eqref{oscP-Meep} can be solved for $P_{n,k}^{m+1}$ in order to update $\vec{P}_n$ from the current ($\vec{P}_n^m$) and previous ($\vec{P}_n^{m-1}$) polarization timesteps, along with the current electric field ($\vec{E}^m$) and population densities ($\vec{N}^m$).

Recall that \emph{Meep} divides the simulation domain into rectilinear ``chunks'' in order to divide the simulation over multiple processors and other reasons~\cite{oskooi_meep:_2010}.  The fact that the population densities $\vec{N}$ must be averaged over neighboring voxels in order to obtain the value at the correct Yee grid point for \eqref{oscP-Meep} means that, in a parallel simulation, the population densities must be communicated between chunks at their boundaries before each timestep.  Similarly, the polarization fields must be communicated along the boundaries (along with the electric fields~\cite{oskooi_meep:_2010}) to perform the averaging in \eqref{oscN-Meep}. This synchronization step allows chunks to use the current populations and polarizations from neighboring chunks in order to update polarizations and populations, respectively, for points on the chunk boundaries.

\subsection{Density matrix model}

Rather than assuming that the dynamics of the bound electrons of the saturable media can be modeled as a damped
harmonic oscillator, it is instead possible to derive this directly beginning from a quantum mechanical model
of the evolution of the saturable medium \cite{boyd_nonlinear_2008,Huang2006,Chang2004},
\begin{equation}
  \partial_t \hat{\rho} = \frac{-i}{\hbar} \left[ \hat{H}_0 + \hat{H}_I, \hat{\rho} \right]. \label{eq:rho}
\end{equation}
Here, $\hat{\rho}$ is the density matrix of an individual atom/molecule of the medium, whose unperturbed Hamiltonian
is $\hat{H}_0$, such that $\hat{H}_0|\psi_i \rangle = E_i | \psi_i \rangle$, and the interaction Hamiltonian from
the incident electric field is $\hat{H}_I = e \hat{\vec{x}} \cdot \vec{E}(\vec{x},t)$.
The evolution for individual density matrix elements in \eqref{rho} can be rewritten and simplified as
\begin{equation}
\partial_t \rho_{ij} = -i\omega_{ij} \rho_{ij} - \frac{i}{\hbar} \sum_k^M \left(\boldsymbol{\theta}_{ik} \rho_{kj} - \rho_{ik} \boldsymbol{\theta}_{kj} \right) \cdot \vec{E}(\vec{x}, t), \label{eq:rho2}
\end{equation}
in which we are using $\boldsymbol{\theta}_{ij} = e \langle \psi_i | \hat{\vec{x}} | \psi_j \rangle$.

The density matrices of
individual atoms/molecules can then be linked to the total macroscopic polarization field as
\begin{equation}
\sum_n \vec{P}_n(\vec{x},t) = -\sum_{\alpha} \delta(\vec{x} - \vec{x}^{(\alpha)}) \textrm{Tr}[\hat{\rho}^{(\alpha)} e \hat{\vec{x}}_\alpha],
\end{equation}
in which $\vec{x}^{(\alpha)}$ and $\hat{\rho}^{(\alpha)}$ are the position and density matrix of atom $\alpha$.
It is also convenient to define the positive frequency polarization component of an individual transition,
\begin{equation}
  \vec{p}_n^+(\vec{x},t) = -\sum_{\alpha} \delta(\vec{x} - \vec{x}^{(\alpha)}) \rho_{ij}^{(\alpha)} \boldsymbol{\theta}_{ji} \equiv -\sum_{\alpha} \delta(\vec{x} - \vec{x}^{(\alpha)}) \rho_{ij}^{(\alpha)} \boldsymbol{\theta}_{n}^*,
\end{equation}
which is related to the classical polarization fields as $\vec{P}_n = 2 \re[\vec{p}_n^+]$,
as well as
\begin{equation}
  N_i(\vec{x},t) = \sum_{\alpha} \delta(\vec{x} - \vec{x}^{(\alpha)}) \rho_{ii}^{(\alpha)}(t).
\end{equation}
Given these definitions, we can rewrite \eqref{rho2} specifically for the occupations,
\begin{equation}
  \partial_t N_i = \sum_{j\ne i}^M \Gamma_{ji}N_j - \sum_{j \ne i}^M \Gamma_{ij} N_i
  + \frac{2}{i\hbar} \sum_n^{N_T} \Xi_{i,n} \im[\vec{p}_n^+] \cdot \vec{E}, \label{eq:rhoD}
\end{equation}
and the polarization components,
\begin{equation}
\partial_t \vec{p}_n^+(\vec{x},t) = -\left(\frac{\gamma_{n}}{2} + i \omega_{n}\right) \vec{p}_n^+ - \frac{i (N_i - N_j)}{\hbar} \boldsymbol{\theta}_n^* \left(\boldsymbol{\theta}_n \cdot \vec{E} \right). \label{eq:rhoP}
\end{equation}
Here, we have added the phenomenological dephasing rates of the transition, $\gamma_n$, as well as the pumping and decay
rates between the different electronic states, $\Gamma_{ij}$.

To make the final connection to the classical oscillator model, \eqrefs{oscP}{oscN}, 
first note that from \eqref{rhoP} and the definition of the classical polarization field,
we can write
\begin{align}
  \partial_t \vec{P}_n &= 2 \partial_t \re[\vec{p}_n^+] \notag \\
  &= 2\left(\omega_n \im[\vec{p}_n^+] - \frac{\gamma_n}{2} \re[\vec{p}_n^+]\right) \label{eq:dtp} \\
  \frac{1}{\omega_n}\left(\partial_t + \frac{\gamma_n}{2} \right)\vec{P}_n &= 2 \im[\vec{p}_n^+],
\end{align}
which, upon substitution into \eqref{rhoD}, completes the derivation of \eqref{oscN}.
To derive \eqref{oscP}, we take a second time derivative of \eqref{dtp},
\begin{align}
  \partial_t^2 \vec{P}_n &= \partial_t \left(2 \omega_n \im[\vec{p}_n^+] - \frac{\gamma_n}{2} \vec{P}_n\right) \notag \\
  &= - \frac{\gamma_n}{2} \partial_t \vec{P}_n + 2 \omega_n \left(-\frac{\gamma_n}{2} \im[\vec{p}_n^+] - \omega_n \re[\vec{p}_n^+] - \frac{N_i - N_j}{\hbar} \boldsymbol{\theta}_n^* \left(\boldsymbol{\theta}_n \cdot \vec{E} \right) \right) \notag \\
  &= - \frac{\gamma_n}{2} \partial_t \vec{P}_n + \left( -\frac{\gamma_n}{2} \left(\partial_t + \frac{\gamma_n}{2} \right)\vec{P}_n - \omega_n^2 \vec{P}_n - \frac{2\omega_n(N_i - N_j)}{\hbar} \boldsymbol{\theta}_n^* \left(\boldsymbol{\theta}_n \cdot \vec{E} \right) \right) \notag \\
  &= -\gamma_n\partial_t \vec{P}_n - \left(\omega_n^2 + \left(\frac{\gamma_n}{2}\right)^2 \right) \vec{P}_n - \frac{2\omega_n(N_i - N_j)}{\hbar} \boldsymbol{\theta}_n^* \left(\boldsymbol{\theta}_n \cdot \vec{E} \right),
\end{align}
which is the result.
As can be seen, the ``extra'' terms discussed at the end of Sec.\ \ref{sec:osc}  appear naturally in deriving
the classical oscillator equations from a microscopic theory. As such, the implementation of saturable media
in \emph{Meep} retains these extra terms to ease comparison against other theories which may possess them.

\section{Natural units of saturable media}

If the saturable medium only possesses a single relevant radiative transition, i.e.\ $N=1$, it is possible to rewrite
\eqrefs{oscP}{oscN} in a dimensionless form. To do so, we first identify the relevant time scale
which dictates the dynamics of the non-radiative decay rates, $\Gamma_{ij}$, and denote this time scale as $\Gamma_{\textrm{ts}}$.
Using this time scale, the natural units of the fields and populations are
\begin{align}
  &\vec{E}_{\textrm{NU}} = \frac{|\theta|}{\hbar \sqrt{\Gamma_{\textrm{ts}} \gamma/2}} \vec{E}, \;\;\;\;\;\;\; \textrm{(Same relation for }\vec{P}\textrm{)} \notag \\
  &N_{i,\textrm{NU}} = \frac{2|\theta|^2}{\hbar \gamma} N_{i}, \notag
\end{align}
and the classical oscillator equations can be written as
\begin{align}
  &\frac{d^2\vec{P}}{dt^2} + \gamma \frac{d\vec{P}}{dt} + \left(\omega_a^2 + \left(\frac{\gamma}{2} \right)^2 \right) \vec{P} = -\omega_a \gamma \Delta N(\vec{x},t) \vec{E}(\vec{x},t), \\
  &\left(\frac{1}{\Gamma_{\textrm{ts}}}\right)\frac{\partial N_i(\vec{x})}{\partial t} = - \sum_j \frac{\Gamma_{ij}}{\Gamma_{\textrm{ts}}} N_i(\vec{x}) + \sum_j \frac{\Gamma_{ji}}{\Gamma_{\textrm{ts}}} N_j(\vec{x}) + \Xi_{i} \vec{E}(\vec{x},t) \cdot \left( \left(\frac{1}{\omega_a}\right)\frac{\partial }{\partial t} + \frac{\gamma}{2\omega_a} \right) \vec{P}(\vec{x},t).
\end{align}
In these equations, we have dropped the subscript $n$ as there is only a single radiative transition everywhere except $\omega_n \rightarrow \omega_a$, which
is denoted as the `atomic' frequency to avoid confusion with the frequency of the fields.

In the special case that a two-level saturable medium is being used, there is a preferred choice of time scale which
can be derived by noting that the total occupancy density in both levels must sum to the total density of atoms/molecules \cite{cerjan_steady-state_2012},
such that
\begin{equation}
  \Gamma_{\textrm{ts}} = \Gamma_{12} + \Gamma_{21},
\end{equation}
and corresponds to the rate at which the inversion, $N_2 - N_1$, decays to its equilibrium value in the absence of any fields.

\section{Validation \label{sec:comp}}

\begin{figure}
  \centering
  \includegraphics[width=0.85\textwidth]{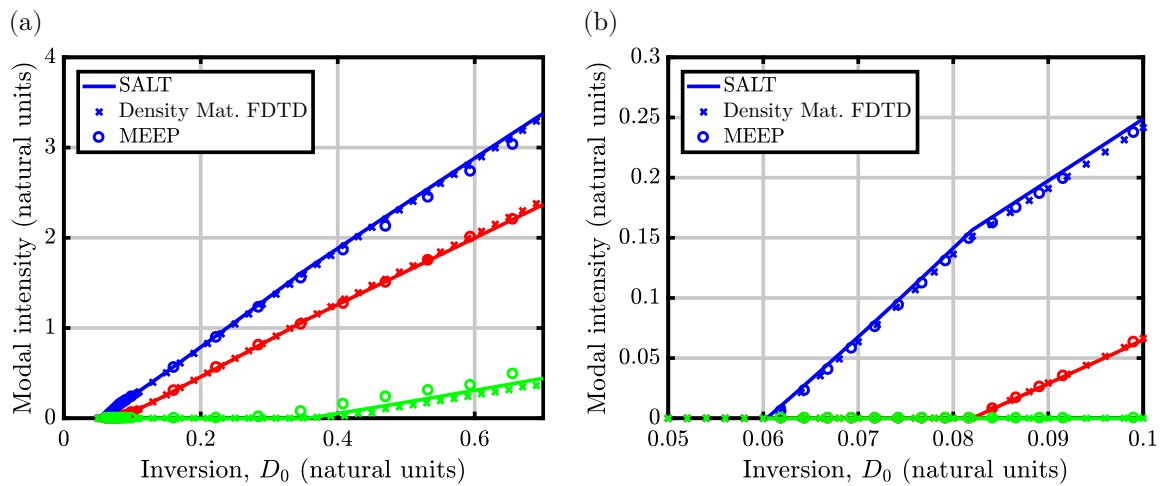}
  \caption{(a) Comparison between the modal intensities found from FDTD simulations using \emph{Meep}, the density matrix equations,
    and SALT through three lasing mode thresholds.
    The system has $\omega_a = 40/(2\pi)$, $\gamma/2 = 4/(2\pi)$, $\Gamma_{21} = 0.005$,
    $\varepsilon = 1.5^2$, and $L = 1$.
    (b) Zoom-in on the region near the first two lasing thresholds.
    \label{fig:1}}
\end{figure}

To verify that saturable media are correctly implemented in \emph{Meep}, we simulate lasing in a one-dimensional Fabry-P{\' e}rot cavity
and compare these results against an earlier FDTD implementation~\cite{cerjan_steady-state_2015} of the density-matrix equations, (\ref{eq:rhoD}) and (\ref{eq:rhoP}),
as well as a nonlinear frequency-domain algorithm for the steady-state lasing solution, SALT \cite{tureci_self-consistent_2006,tureci_strong_2008,ge_steady-state_2010,esterhazy_scalable_2014}.
The cavity consists of a dielectric slab, $n = 1.5$, with a perfectly-reflecting mirror on one end and an open facet out of which light can escape
on the other end. Confinement of the field inside the cavity is strictly due to Fresnel reflection at the interface. This dielectric cavity is filled
with a saturable two-level medium, in which the lower electronic state is being pumped to the upper state at a faster rate than the upper state
non-radiatively relaxes to the lower state, $\Gamma_{12} > \Gamma_{21}$. In dimensionless \emph{Meep} units ($2\pi c/a$), we take the transition frequency to be
$\omega_a = 40/(2\pi)$, the dephasing rate to be $\gamma/2 = 4/(2\pi)$ (also in units of $2\pi c/a$), and the non-radiative decay rate to be $\Gamma_{21} = 0.005$
(in units of $c/a$, \textit{not} $2\pi c/a$),
and assign the cavity a length of $L = 1$. $\Gamma_{12}$ is varied to change the strength of the gain in the system, while the total density of saturable
atoms, $N_{\textrm{atom}}$, is fixed.
In real units, if the lasing wavelength
is $\lambda = \SI{900}{\nano\meter}$, that corresponds to a cavity length of approximately $L \approx \SI{6}{\micro\meter}$, which is an unphysically
short cavity, but still useful for numerical comparisons. As a final note for this comparison, we use the inversion in the absence of the electric field,
$D_0$, as the effective pump parameter, which is defined in terms of the pumping and decay rates as
\begin{equation}
  D_0 = \frac{\Gamma_{12} - \Gamma_{21}}{\Gamma_{12} + \Gamma_{21}} N_{\textrm{atom}},
\end{equation}
in which $N_{\textrm{atom}}$ is the density of saturable atoms/molecules, $N_{\textrm{atom}} = N_1 + N_2$.
As can be seen in Fig.\ \ref{fig:1}, \emph{Meep} simulations which are run sufficiently long for the system to ready its steady state agree with similar
results from an independent FDTD based on the density matrix equations,
as well as the steady-state lasing solution from SALT.

\begin{figure}
  \centering
  \includegraphics[width=0.85\textwidth]{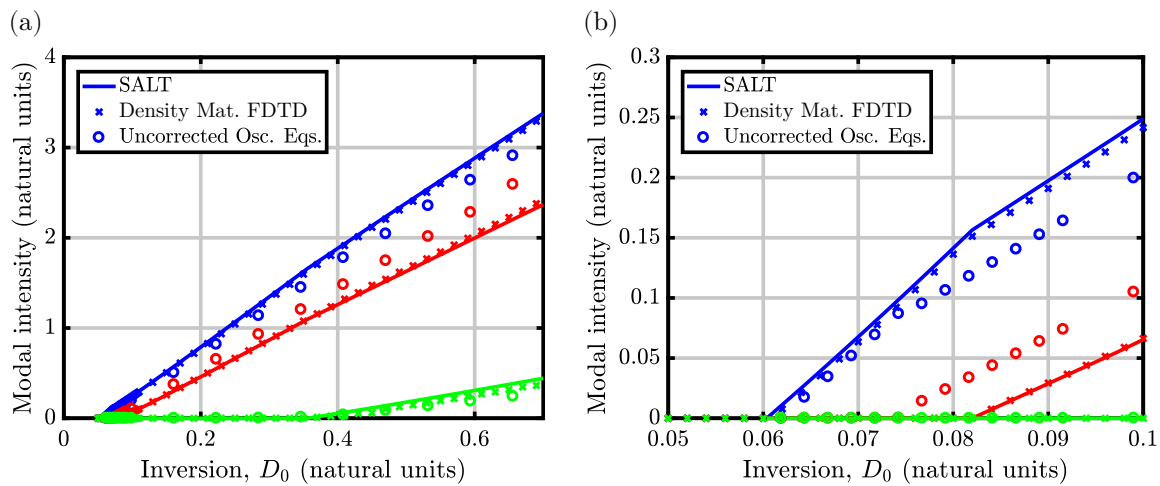}
  \caption{(a) Comparison between the modal intensities found from FDTD simulations using \emph{Meep} with the `uncorrected' classical oscillator equations,
    the density matrix equations, and SALT through three lasing mode thresholds.
    The system has $\omega_a = 40/(2\pi)$, $\gamma/2 = 4/(2\pi)$, $\Gamma_{21} = 0.005$,
    $\varepsilon = 1.5^2$, and $L = 1$.
    (b) Zoom-in on the region near the first two lasing thresholds.
    \label{fig:2}}
\end{figure}

In addition, it is worth emphasizing here that the two terms which are typically approximated to zero in the classical oscillator equations,
following the discussion from the end of Sec.\ \ref{sec:osc}, can play a significant role in understanding the details of the laser's operation.
Shown in Fig.\ \ref{fig:2} are simulations of the same laser system, but without these two extra terms in \eqrefs{oscP}{oscN},
and as can be seen, there are significant deviations in the lasing thresholds and modal intensities.

\section{Using saturable media in \emph{Meep}}

In this section, we discuss how the parameters discussed in Sec.\ \ref{sec:osc} map to \emph{Meep}'s Python API.
In \emph{Meep}, saturable media are defined using the \texttt{MultilevelAtom} class, which is a subclass of \texttt{E\_susceptibilities}.
There are two separate objects that one must specify to properly initialize \texttt{MultilevelAtom} -- a set of radiative and
non-radiative transitions that are both specified using \texttt{Transition}, and a set of initial populations of each of the electronic
levels. The latter is straightforward, and is a list of the occupancies of the electronic states, $[N_1, N_2,\ldots]$, at $t = 0$.
To define a non-radiative transition, i.e.\ an element $\Gamma_{ij}$, the the two levels must be specified, as well as the transition rate,
\begin{lstlisting}
     meep.Transition(from_level=$i$, to_level=$j$, transition_rate=$\Gamma_{ij}$)
\end{lstlisting}
It is important to note that these non-radiative rates are specified in units of $c/a$, \textit{not} the usual \emph{Meep} frequency units of $2\pi c/a$.
If $i>j$, this represents a non-radiative decay rate, whereas if $i<j$ this represents a pumping rate, but both can be specified using 
\texttt{transition\_rate}. To instead specify a radiative transition between two levels, which will implicitly initialize a corresponding 
non-linear polarization field $\mathbf{P}_n$, one must specify all of the necessary criteria for this transition,
\begin{minipage}{\linewidth}
\begin{lstlisting}
     meep.Transition(from_level=$i$, to_level=$j$, frequency=$\omega_{n}$,
                     gamma=$\gamma_n$, sigma_diag=$\textrm{diag}[\bar{\sigma}_n]$)
\end{lstlisting}
\end{minipage}
In this case, it does not matter in what order you specify $i$ and $j$, and both $\omega_n$ and $\gamma_n$ are specified in 
\emph{Meep}'s frequency units of $2\pi c/a$. Here, $\textrm{diag}[\bar{\sigma}_n]$ is specified as a \texttt{meep.Vector3}.
If you have both a radiative and non-radiative transition between two levels, which is common
because the upper level can be metastable, but not completely stable (as that would be a ground state), you can specify both processes
in a single instance of \texttt{Transition},
\begin{lstlisting}
     meep.Transition(from_level=$i$, to_level=$j$, transition_rate=$\Gamma_{ij}$
                     frequency=$\omega_{n}$, gamma=$\gamma_n$, sigma_diag=$\textrm{diag}[\bar{\sigma}_n]$)
\end{lstlisting}
in which case the ordering of $i$ and $j$ does matter. 
At present, off-diagonal elements in  $\bar{\sigma}_n$ are not supported, as discussed in Sec.\ \ref{sec:impl}.

Then, given a list of transitions, as well as a list of initial populations, one can define a multilevel atom susceptibility as
\begin{lstlisting}
     ml_atom = meep.MultilevelAtom(transitions=[$\textrm{list of transitions}$], 
                                   initial_populations=[$N_1(t=0)$,$N_2(t=0)$,...])
\end{lstlisting}
which can now be added to any specification of a geometric object's material (or to the background medium of a simulation),
as
\begin{lstlisting}
     material = meep.Medium(index=$n_{\textrm{cav}}$, E_susceptibilities=[ml_atom])
\end{lstlisting}
Here, $n_{\textrm{cav}}$ is the index of the linear response of the medium, independent of the non-linear saturable medium.

Given such a saturable-gain medium, in order to observe lasing one must initialize the electromagnetic field to a nonzero value
(either by a short-lived current source or using the \texttt{initialize\_field} function).  Otherwise, there is no field to amplify into the
lasing mode---the coupling between the electronic populations and the electromagnetic field in \eqrefs{oscP}{oscN} is zero when $\vec{E}=0$.
(In a physical system, nonzero fields are created by thermodynamic fluctuations.)

For the Python script used to generate the results in Sec.\ \ref{sec:comp}, see the \href{https://meep.readthedocs.io/en/latest/Python_Tutorials/Multilevel_Atomic_Susceptibility/}{tutorial example}
in the \emph{Meep} user manual.

\section{Conclusion}

In this technical note, we have described the physical model and numerical implementation of saturable media in \emph{Meep}. Feature requests, bug reports,
and suggestions for general improvements are welcome and can be made through either the user mailing list \href{mailto:meep-discuss@ab-initio.mit.edu}{meep-discuss@ab-initio.mit.edu}
or as a Github issue via the source repository \url{https://github.com/NanoComp/meep}.

\section*{Acknowledgements}

This work was supported in part by the U.~S.~Army Research Office through the Institute for Soldier Nanotechnologies (award~W911NF-18-2-0048)
as well as by the National Science Foundation (NSF) via Small Business Innovation Research (SBIR) Phase I and II awards 1647206 and 1758596.
A.C.\ acknowledges the support of the US Office of Naval Research (ONR) Multidisciplinary University Research Initiative (MURI) grant N00014-20-1-2325
on Robust Photonic Materials with High-Order Topological Protection.


%

\end{document}